\documentclass[aps,prb,preprint,superscriptaddress]{revtex4-2}
\usepackage{CJK}
\usepackage{graphicx}
\usepackage{amsmath}
\usepackage{amssymb}
\usepackage {color} 
\usepackage{kotex}
\UseRawInputEncoding
\begin{document}
\begin{CJK*}{GB}{} 


\title{Maximum plasmon thermal conductivity of a thin metal film}


\author{Kuk Hyun Yun}
\thanks{These two authors contributed equally}
\author{Dong-min Kim}
\thanks{These two authors contributed equally}
\affiliation{Department of Mechanical Engineering, Korea Advanced Institute of Science and Technology, Daejeon 34141, South Korea}
\affiliation{Center for Extreme Thermal Physics and Manufacturing, Korea Advanced Institute of Science and Technology, Daejeon 34141, South Korea}

\author{Bong Jae Lee}
\email{bongjae.lee@kaist.ac.kr}
\affiliation{Department of Mechanical Engineering, Korea Advanced Institute of Science and Technology, Daejeon 34141, South Korea}
\affiliation{Center for Extreme Thermal Physics and Manufacturing, Korea Advanced Institute of Science and Technology, Daejeon 34141, South Korea}


\date{\today}
\maketitle
\end{CJK*}

\section*{Abstract}
Due to their extremely long propagation lengths compared to the wavelengths, surface plasmon polaritons (SPPs) have been considered as a key in enhancing thermal conductivity in thin metal films. This study explores the conditions at which the plasmon thermal conductivity is maximized, considering the thickness-dependent metal permittivity. We derived the analytical solutions for the plasmon thermal conductivity in both the thin-film and thick-film limits to analyze the effect of the permittivities of metals and substrates. From the analytical solutions of plasmon thermal conductivity, we deduced that the plasmon thermal conductivity is proportional to the electron thermal conductivity based on the Wiedemann-Franz law. Additionally, we analyzed the conditions where the enhancement ratio of the thermal conductivity via SPPs is maximized. Metals with high plasma frequency and low damping coefficient are desirable for achieving the maximum plasmon thermal conductivity as well as the maximum enhancement ratio of thermal conductivity among metals. Significantly, 10-cm-long and 14-nm-thick Al film demonstrates most superior in-plane heat transfer via SPPs, showing a 53.5\% enhancement in thermal conductivity compared to its electron thermal counterpart on a lossless glass substrate.




\section{Introduction}
When the characteristic length of thin films or the diameter of nanowires becomes comparable to the mean free path of basic energy carriers (e.g., phonons and electrons), boundary scattering becomes increasingly significant \cite{chen2005nanoscale, zhang2007nano}. This effect causes a reduction in the effective mean free path of energy carriers compared to their bulk counterparts, leading to a decrease in thermal conductivity \cite{feng2009prediction, jeong2012thermal, wang2014computational}. This classical size effect of nanostructures can cause significant issues, such as performance degradation and reliability reduction in modern semiconductor devices, intensifying the focus on thermal management challenges \cite{moore2014emerging, zheng2021advances}. In response to these challenges, most research has been directed toward replacing conventional materials in modern devices with those having higher thermal conductivities \cite{zhou2017thermal}. Recently, two-dimensional (2-D) materials with pronounced anisotropic thermal conductivity have drawn attention as heat spreaders \cite{yan2012graphene, bae2014graphene, jeon2021suppression}. They possess very high in-plane thermal conductivities ranging from hundreds to thousands W/m$\cdot$K, while their cross-plane thermal conductivities remain significantly lower. With their tens of nanometer thickness, such 2-D materials enable concentrated in-plane heat conduction, efficiently diffusing heat away from the hot spots. However, as of now, the commercialization of 2-D material heat spreader is challenging due to the challenges in fabrication and assembly on the devices \cite{song2018two}.

Recently, it has been reported that surface waves, such as surface phonon polaritons (SPhPs) and surface plasmon polaritons (SPPs) in nanoscale thin films, can compensate for the classical size effect via additional heat conduction channels \cite{chen2005surface, tranchant2019two, wu2020enhanced, wu2022observation, kim2023boosting, pan2023remarkable, pei2023low}. In fact, surface waves can propagate over distances longer than a centimeter, significantly enhancing thermal conductivity \cite{ordonez2013anomalous, ordonez2014thermal, ordonez2023plasmon}. Therefore, nanoscale metal films can maintain the high thermal conductivity comparable to their bulk values due to additional in-plane heat transfer via SPPs, suggesting that they can serve as heat spreaders similar to 2-D materials. Furthermore, unlike 2-D materials, thin metal films can be easily deposited on devices through a microelectromechanical systems process. Recently, \citet{ordonez2023plasmon} theoretically demonstrated an increase of 25\% in the thermal conductivity of a metal due to long-range SPPs for a 1-cm-long gold film deposited on a Si substrate. However, this increase rate is overestimated to some extent because the plasmon thermal conductivity does not take into account the thickness-dependent metal permittivity. More recently, \citet{kim2023plasmon} comprehensively accounted for the size effect on the permittivity of metals, calculating and experimentally verifying the plasmon thermal conductivity in Au and Ag thin films. They showed that for 5-cm-long Au and Ag thin films, the plasmon thermal conductivity reaches about 20\% of its electron thermal conductivity.

However, it remains unclear yet what optical properties of metals determine the plasmon thermal conductivity and what the optimal conditions are for maximizing it. To analyze the effect of the permittivities of metal film and dielectrics (i.e., substrate or superstrate) on plasmon thermal conductivity, it is necessary to analytically express the plasmon thermal conductivity as a function of the permittivities. This is because the properties of SPPs, such as the in-plane wavevector and propagation length, are determined by the permittivities of media. In other words, the material combinations of the metal film and dielectrics are a primary factor in determining the magnitude of plasmon thermal conductivity. But, the analytical solution of dispersion relation of SPPs exhibits intricate dependence with the permittivity of media \cite{ordonez2023plasmon}, making it difficult to analyze its effect on plasmon thermal conductivity.

In this study, we present criteria that maximize the plasmon thermal conductivity of a thin metal film. This is achieved by analytically deriving expressions for the plasmon thermal conductivity as a function of the film thickness and the permittivity of a metal film and a substrate, considering the thickness-dependent metal permittivity using a modified Drude model. These analytic expressions demonstrate a monotonic dependence of the plasma frequency and damping coefficient on the plasmon thermal conductivity, through approximation with respect to the film thickness. We also investigated the ratio of the plasmon thermal conductivity to the electron thermal conductivity in a metal film. Finally, we proposed which of the existing metals are most desirable for the SPP-mediated in-plane heat transfer.


\section{Theoretical model}

Consider the in-plane heat transfer via SPPs along an infinitely long metal film surrounded by dielectrics (i.e., substrate and superstrate). The plasmon thermal conductivity can be calculated from kinetic theory with Boltzmann transport equation (BTE) under relaxation time approximation and diffusion approximation \cite{chen2005surface, tranchant2019two}: 
\begin{equation} \label{Eq:kSPP}
	 k_{\text{SPP}}=\frac{1}{4\pi d_m}\int^{\infty}_{0}\hbar\omega\beta_R\Lambda_{\text{SPP}}\frac{\partial f_0}{\partial T}d\omega,
\end{equation}
where $d_m$ is the thickness of the metal film, $\hbar$ is the Planck constant divided by 2$\pi$, $\omega$ is the angular frequency, $\beta_R$ is the real part of in-plane wavevector of SPP (i.e., $\beta = \beta_R + i\beta_I)$, $\Lambda_{\text{SPP}}$ is the propagation length of SPP, $f_0$ is the Bose-Einstein distribution function, and $T$ is the temperature. For a metal film with finite-length of $L_m$, the effective propagation length derived from the BTE can be used to consider the boundary scattering, i.e., $\Lambda_{\text{eff}} = [1 - 4\psi(0)/(\pi\mu)] \Lambda_{\text{SPP}}$ \cite{guo2021quantum}. Here, $\mu = L_m / \Lambda_{\text{SPP}}$ and $\psi(\xi) = E_5 (\xi) - E_5 (\mu - \xi)$, where $\xi = z /\Lambda_{\text{SPP}}$ and $E_n (x) = \int_{0}^{\pi/2} (\cos \theta)^{n-2} \exp(-x/\cos\theta) d\theta$ with $\theta$ being the polar angle between the SPP propagation direction and global heat transfer direction. To calculate the plasmon thermal conductivity, the in-plane wavevector and propagation length of SPPs should first be obtained by solving the dispersion relation. In this study, the three-layer configuration consisting of a substrate, metal film, and superstrate is considered. For simplicity, the superstrate is set to be air, and the corresponding three-layer dispersion relation for SPPs is given by \cite{yeh2008plasmon}
\begin{equation} \label{Eq:dispersion}
	 \tanh (p_m d_{m}) = -\frac{p_{m}\varepsilon_{m}(p_s\varepsilon_a+p_a\varepsilon_s)}{p_{m}^2\varepsilon_s\varepsilon_a+p_sp_a\varepsilon_{m}^2},
\end{equation}
where the subscripts `$m$', `$s$', and `$a$' represents metal, substrate, and air, respectively. Additionally, $p_n = \sqrt{\beta^2 - \varepsilon_n k_0^2}$ denotes the cross-plane wavevector of SPPs for medium $n = m, s$, or $a$, where $\varepsilon_n$ is the relative permittivity of the corresponding medium, $k_0 = \omega / c_0$ is the wavevector in a vacuum with $c_0$ being the speed of light in a vacuum. In this study, we consider only the metal as a lossy material (i.e., $\varepsilon_m = \varepsilon_R + i\varepsilon_I$), while treating the substrate as a lossless dielectric for simplicity.

It is well known that SPPs at both interfaces of a metal film become decoupled (i.e., SPPs at two interfaces behave independently of each other) as the film thickness increases to the optically thick limit (i.e., thick-film limit) \cite{maier2007plasmonics}. In the thick-film limit, the plasmon thermal conductivity increases as the film thickness decreases, as seen in Eq.\ \eqref{Eq:kSPP}. The plasmon thermal conductivity reaches its peak at the thickness where the decoupling of SPPs begins (i.e., $\text{Re}(p_{m})d_{m}\approx1$) \cite{kim2023plasmon}. When the film thickness is sufficiently thin, the SPPs at both interfaces start to couple, resulting in the thickness-dependent SPP disperion relation. Such intensification of SPPs coupling lead to increased energy losses within the metal film, causing the plasmon thermal conductivity to decrease accordingly \cite{ordonez2023plasmon}. Therefore, to analyze the peak of plasmon thermal conductivity, understanding the behavior of SPPs just before decoupling occurs is crucial.

In the thin-film limit when $|p_{m}|d_{m}\ll1$, the left side of Eq.\ \eqref{Eq:dispersion} can be approximated to $\tanh(p_{m}d_{m})\approx p_{m}d_{m}$. Therefore, the dispersion relation becomes
\begin{equation} \label{Eq:approxi dispersion 1}
	 d_{m}(p_{m}^{2}\varepsilon_{s}\varepsilon_{a}+p_{s}p_{a}\varepsilon_{m}^{2})=-\varepsilon_{m}(p_{s}\varepsilon_{a}+p_{a}\varepsilon_{s}),
\end{equation}
which is an implicit equation for $\beta$, making it difficult to obtain an analytic solution. If the cross-plane wavevector for air is expressed as $p_{a}=\sqrt{\alpha}=\sqrt{\beta^2-\varepsilon_{a}k_0^2}$, then the cross-plane wavevector for the substrate can be expressed as $p_{s}=\sqrt{\alpha-(\varepsilon_s-\varepsilon_{a}) k_0^2}$. Additionally, if the magnitude of permittivity of the metal is much greater than that of the substrate and air (i.e., $|\varepsilon_{m}|\gg|\varepsilon_{s}|,|\varepsilon_{a}|$), the cross-plane wavevector for metal becomes $p_{m}\approx k_{0}\sqrt{-\varepsilon_{m}}$ \cite{ordonez2023plasmon}. Consequently, with those approximations in the thin-film limit, the implicit dispersion relation in Eq.\ \eqref{Eq:approxi dispersion 1} can be written to the explicit form for $\alpha$ as
\begin{equation} \label{Eq:approxi dispersion 2}
	 \varepsilon_{m}^{4}d_{m}^{4}\alpha^{4}-2(\varepsilon_{s}-\varepsilon_{a})\varepsilon_{m}^{4}k_0^{2}d_{m}^{4}\alpha^{3}+(\varepsilon_{s}-\varepsilon_{a})^{2}\varepsilon_{m}^{4}k_0^{4}d_{m}^{4}\alpha^{2}-(\varepsilon_{s}-\varepsilon_{a})^{2}\varepsilon_{m}^{2}\varepsilon_{a}^{2}k_0^{4}d_{m}^{2}\alpha+(\varepsilon_{s}-\varepsilon_{a})^{2}\varepsilon_{a}^{4}k_0^{4}=0.
\end{equation}
The in-plane wavevector of SPPs can be easily determined from the explicit form of a fourth-order polynomial of $\alpha$. For long-range SPPs with propagation lengths exceeding centimeters, the imaginary part of the in-plane wavevector should be considerably small ($\Lambda_\text{SPP}=1/(2\beta_I)$), and the real part of the in-plane wavevector is nearly identical to the light line (i.e., $\beta_R\approx k_0\sqrt{\varepsilon_{a}}$). This implies that the real part of the in-plane wavevector is significantly larger than its imaginary part ($|\beta_R|\gg|\beta_I|$), allowing the magnitude of the in-plane wavevector to be assumed equal to its real part. With this assumption, the magnitude of $\sqrt{\alpha}$ becomes very small compared to the light lines (i.e., $\sqrt{\alpha}=\sqrt{\beta^{2}-\varepsilon_{a}k_0^{2}}\ll k_0\sqrt{\varepsilon_{a}}$). Therefore, $\alpha\ll(\varepsilon_{s}-\varepsilon_{a})k_0^{2}$ is satisfied except in cases when the substrate is identical to the superstrate. As a result, Eq.\ \eqref{Eq:approxi dispersion 2} can be further approximated to a second-order polynomial of $\alpha$. The solution of second-order polynomial, the cross-plane wavevector for air (i.e., superstrate), becomes
\begin{equation} \label{Eq:approxi cross}
	 \sqrt{\alpha}=p_{a}=\frac{\varepsilon_{a}}{\varepsilon_{m}d_{m}}.
\end{equation}
Equation\ \eqref{Eq:approxi cross} indicates that the penetration depth of the SPPs into the air (i.e., $\delta_{a}=1/(2\text{Re}(p_{a}))$) increases as the film thickness increases. The frequency-dependent permittivity of metal can be described using the Drude model
\begin{equation} \label{Eq:drude}
	 \varepsilon_{m}(\omega,d_{m})=1-\frac{\omega_p^2}{\omega^2+i\Gamma(d_{m})\omega},
\end{equation}
where $\omega_{p}$ is the plasma frequency and $\Gamma$ is the thickness-dependent damping coefficient, which signifies the collision of electrons with phonon, grain boundaries, and film boundaries \cite{zhang2007nano}. In other words, the damping coefficient, as the inverse of the free electron's relaxation time, enables the consideration of boundary scattering. By applying Matthiessen's rule to account for boundary scattering, the damping coefficient can be related to its bulk value ($\Gamma_{\infty}$), as $\Gamma=\Gamma_{\infty}+v_{f}/d_{m}$, where $v_{f}$ is the Fermi velocity \cite{otanicar2010nanofluid, lee2012radiative}. As the film thickness decreases, the damping coefficient increases, leading to more losses in the metal. If the angular frequency of SPPs is much lower than the plasma frequency and is much higher than the damping coefficient (i.e., $\Gamma \ll \omega \ll \omega_{p}$), the real and imaginary parts of the permittivity can be approximated to $\varepsilon_{R}=-\omega_{p}^{2}/\omega^{2}$ and $\varepsilon_{I}=\omega_{p}^{2}\Gamma/\omega^{3}$, respectively. By substituting the metal permittivity derived from Drude model into Eq.\ \eqref{Eq:approxi cross}, the analytical solution of the dispersion relation in the thin-film limit can be expressed by
\begin{subequations}
\label{Eq:approxi Beta}
\begin{align}
    \label{Eq:approxi betar}
    \beta_{R}=k_0\sqrt{\varepsilon_{a}}+\frac{\varepsilon_{a}^{3/2}}{2k_0 d_{m}^{2}} \left (\frac{\omega}{\omega_{p}} \right)^{4},\\
    \label{Eq:approxi betai}
    \beta_{I}=\frac{c_0\varepsilon_{a}^{3/2}}{d_{m}^{2}}\frac{\Gamma}{\omega_{p}^{2}} \left (\frac{\omega}{\omega_{p}} \right)^{2}.
\end{align}
\end{subequations}
Equation\ \eqref{Eq:approxi betar} indicates that in the thin-film limit, the SPP dispersion curve moves further away from the light line as the film becomes thinner. Note that if the angular frequency is much lower than the plasma frequency, the dispersion curve becomes simply the light line (i.e., $\beta_R\approx k_0\sqrt{\varepsilon_{a}}$). Concurrently, the imaginary part of the in-plane wavevector increases as the film thickness decreases, which reduces the propagation length of the SPPs. In the thinner film, the SPPs at both interfaces become coupled, resulting in increased energy dissipation within the metal film. This implies that the energy losses from coupled SPPs lead to a reduction in their propagation lengths. Such a decrease in SPP propagation lengths is responsible for the pronounced reduction in the plasmon thermal conductivity.

Equations\ \eqref{Eq:approxi dispersion 2},\ \eqref{Eq:approxi cross}, and\ \eqref{Eq:approxi Beta} represent the solutions about the dispersion relation of SPPs propagating along the air/metal interface. Due to the symmetry of dispersion relation for air and substrate (see Eq.\ \eqref{Eq:dispersion}), the in-plane wavevector for the substrate can be obtained simply by replacing the subscript `$a$' with `$s$'. By substituting Eqs.\ \eqref{Eq:approxi betar} and \eqref{Eq:approxi betai} into Eq.\ \eqref{Eq:kSPP}, the plasmon thermal conductivity in the thin-film limit can be written as
\begin{equation} \label{Eq:kSPPthin}
	 k_{\text{SPP,thin}} = \sum_{n=s,a} \frac{\hbar d_m}{8\pi\epsilon_n c_0^2} \frac{\omega_p^4}{\Gamma} \int_0^\infty \frac{\partial f_0}{\partial T} d\omega.
\end{equation}
The above equation suggests that in the thin-film limit, the plasmon thermal conductivity increases as the film thickness increases (i.e., $k_{\text{SPP,thin}} \propto d_{m}$). Note also that the damping coefficient decreases as the film thickness increases, leading to an increase in the plasmon thermal conductivity. Equation\ \eqref{Eq:kSPPthin} also suggests that $k_{\text{SPP,thin}}$ increases for metals with a higher plasma frequency and a lower damping coefficient, and it also increases with lower permittivity of the substrate. For a given substrate and superstrate, the gradient of plasmon thermal conductivity with respect to film thickness is solely determined by the permittivity of metal. That is, metals with a high plasma frequency and a low damping coefficient represent a steeper increase in the plasmon thermal conductivity as $d_{m}$ increases. For instance, for Au ($\omega_{p}=64660$ cm$^{-1}$, $\Gamma_{\infty}=252$ cm$^{-1}$, and $v_{f}=13.82\times10^{5}$ m/s) and Ag ($\omega_{p}=72071$ cm$^{-1}$, $\Gamma_{\infty}=145$ cm$^{-1}$, and $v_{f}=14.48\times10^{5}$ m/s) films, Ag has a steeper increase in the plasmon thermal conductivity in the thin-film limit than that of Au \cite{kim2023plasmon}.

On the other hand, when a metal film thickness becomes optically thick, the SPPs at both interfaces of the metal film become completely decoupled and the corresponding SPP dispersion curve is independent of film thickness. The plasmon thermal conductivity using the analytical solution for the dispersion relation of SPPs in the thick-film limit is given by \cite{ordonez2023plasmon}
\begin{equation} \label{Eq:kSPPthick}
	 k_{\text{SPP,thick}} = \sum_{n=s,a} \frac{\hbar}{4\pi d_{m} \varepsilon_n} \frac{\omega_p^2}{\Gamma_\infty} \int_0^\infty \frac{\partial f_0}{\partial T} d\omega.
\end{equation}
In the thick-film limit, $k_{\text{SPP,thick}}$ is inversely proportional to the film thickness, which is the opposite of what is observed in the thin-film limit. However, similarly to the thin-film limit $k_{\text{SPP,thick}}$ increases also for metals with a higher plasma frequency and a lower damping coefficient, and for substrates with lower permittivity. Note that in both limits, the plasmon thermal conductivity has a factor of $\omega_{p}^{2}/(\Gamma\varepsilon_{n})$. Based on the Drude model and Wiedemann-Franz law, the electron thermal conductivity ($k_{e}$) can be expressed by $k_{e}=\varepsilon_0 L T \omega_{p}^{2}/\Gamma$, where $\varepsilon_0$ is the vacuum permittivity and $L$ is the Lorenz number for each metal \cite{avery2015thermal, kittel2018introduction}. This clearly implies that the plasmon thermal conductivity is proportional to the electron thermal conductivity in both limits. Metals with a high plasma frequency have a larger number density of free electrons \cite{maier2007plasmonics}. This larger number density of free electrons facilitates greater energy transfer, resulting in enhanced electrical and thermal conductivities. Likewise, this principle is also applicable to SPPs, which are energy carriers formed by the coupling of free electrons and photons. In other words, an increase in the number density of free electrons leads to the enhancement of energy transfer via SPPs, thereby improving the plasmon thermal conductivity.

\section{Results and discussion}
We intentionally selected five metals whose phonon contribution to the total intrinsic thermal conductivity (i.e., thermal conductivity via electron and phonon) is less than 10\% to further investigate the relationship between the plasmon thermal conductivity and the electron thermal conductivity, as listed in Table\ \ref{table 1}. For each metal, its Drude parameters (i.e., plasma frequency and damping coefficient) for low frequencies were obtained by the least-square fitting of experimental data in Refs.\ \cite{ordal1985optical, ordal1988optical}, and for high frequencies, from tabulated data \cite{palik1998handbook}. In addition, their Fermi velocities were taken from values predicted by the density functional theory \cite{gall2016electron}. The electron thermal conductivity of Ag and Al, predicted using the fitted Drude parameters and the Wiedemann-Franz law, showed good agreement with the literature values \cite{bergman2011fundamentals}. However, for the remaining metals (Cu, Au, and Pt), the calculated values deviate from the measurements by about 30 to 50\%. Such discrepancy can be attributed to the limitations of the Drude model, which simplifies the behavior of electrons in metals to that of free particles and fails to account for complex interactions such as electron-electron interactions \cite{dressel2006verifying}. Note that the main focus of this study lies on analyzing the effect of Drude parameters on the plasmon thermal conductivity. Therefore, in this study, the thermal conductivity of metals is simply considered based on the electron thermal conductivity predicted from the Drude model and the Wiedemann-Franz law.

\begin{table}[t!]
\centering
\caption{This table includes Drude parameters, Fermi velocities, and thermal conductivity for various metals. $k_{intrin}$ denotes the total intrinsic thermal conductivity of metal, $k_{e}$ is the electron thermal conductivity predicted by Wiedemann-Franz law, $k_{ph}$ is the phonon thermal conductivity, and $k_{total} = k_{e} + k_{ph}$ is the total thermal conductivity. $L_0=2.44\times10^{-8}$ W$\cdot\Omega/\text{K}^{2}$ is the Sommerfeld value of Lorenz number ($L$).}
\label{table 1}
\begin{tabular}{@{\hspace{5pt}}c@{\hspace{7pt}}|@{\hspace{5pt}}cc@{\hspace{5pt}}c@{\hspace{5pt}}c@{\hspace{5pt}}c@{\hspace{5pt}}c@{\hspace{5pt}}c}
\hline \hline
Metal & \begin{tabular}[c]{@{}c@{}} $\omega_{p}$ (cm$^{-1}$) \\ \cite{ordal1985optical, ordal1988optical, palik1998handbook} \end{tabular} & \begin{tabular}[c]{@{}c@{}} $\Gamma_{\infty}$ (cm$^{-1}$) \\ \cite{ordal1985optical, ordal1988optical, palik1998handbook} \end{tabular} & \begin{tabular}[c]{@{}c@{}} $v_{f}$ (10$^{5}$ m/s) \\ \cite{gall2016electron} \end{tabular} & \begin{tabular}[c]{@{}c@{}} $L/L_{0}$ \\ \cite{tong2019comprehensive} \end{tabular} & \begin{tabular}[c]{@{}c@{}} $k_{e}$ (W/m$\cdot$K) \\ \end{tabular} & \begin{tabular}[c]{@{}c@{}} $k_{intrin}$ (W/m$\cdot$K) \\ \cite{bergman2011fundamentals} \end{tabular} & \begin{tabular}[c]{@{}c@{}} $k_{ph}/k_{total}$ (\%) \\ \cite{tong2019comprehensive} \end{tabular} \\ \hline \hline
Ag & 72071 & 145 & 14.48 & 0.98 & 428.6 & 429 & 1.25 \\
Cu & 59022 & 76.1 & 11.09 & 0.94 & 525.3 & 401 & 4.60 \\
Au & 64660 & 252 & 13.82 & 1.03 & 208.6 & 317 & 1.01 \\
Al & 96627 & 437 & 15.99 & 0.94 & 245.2 & 237 & 3.71 \\
Pt & 41775 & 614 & 5.2 \cite{dutta2017thickness} & 1.0 & 34.70 & 71.6 & 6.74 \\ \hline \hline
\end{tabular}
\end{table}

Initially, to observe the effect of metal's permittivity on the plasmon thermal conductivity, the superstrate and substrate were fixed as air and glass (amorphous SiO$_{2}$ neglecting losses), respectively. The permittivity of lossless glass was set to be 3.6955 for below 200 Trad/s, which dominantly contributes to the plasmon thermal conductivity \cite{kim2023boosting}. This value is the average obtained from tabulated data for frequency below 200 Trad/s \cite{palik1998handbook}. In the calculation, we fixed the lateral size of the metal film as $L_m=10$ cm. As previously explained in Eq.\ \eqref{Eq:kSPP}, the plasmon thermal conductivity is calculated using the effective propagation length in finite-length metal films. Since the effective propagation length is proportional to the intrinsic propagation length of SPPs, Eqs.\ \eqref{Eq:kSPPthin} and\ \eqref{Eq:kSPPthick} for infinite-length film can adequately explain the plasmon thermal conductivity for finite-length films with respect to the permittivity of the metal. We will address this matter later in the discussion.

\begin{figure}[b!]%
\centering
\includegraphics[width=0.5\textwidth]{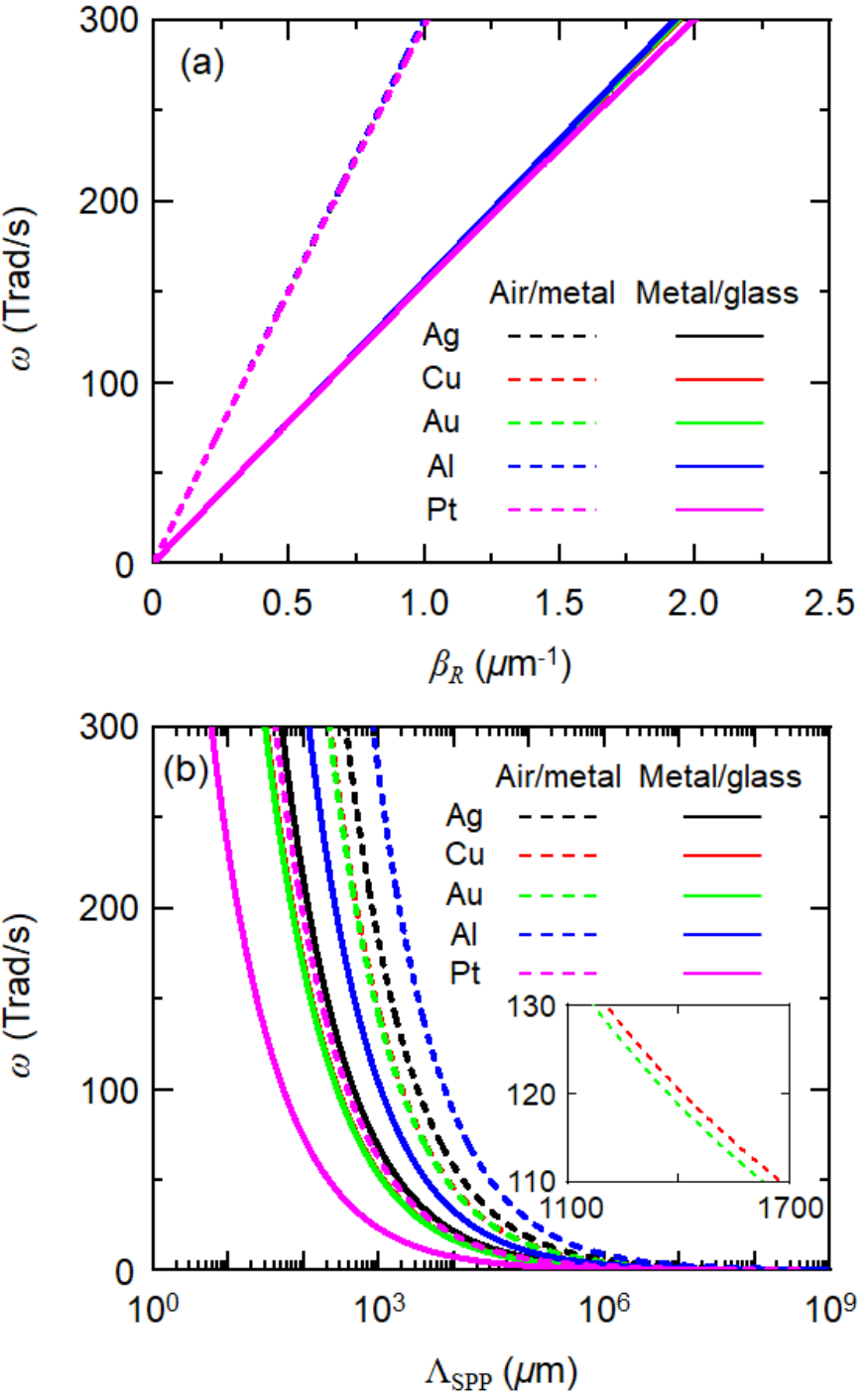}
\caption{(a) In-plane wavevector and (b) propagation length of SPP along the interface of 10-nm-thick metal film supported by lossless glass substrate. The solid lines represent the metal/glass interface, while the dashed lines indicate the air/metal interface.}\label{fig:thin dispersion}
\end{figure}

Figures\ \ref{fig:thin dispersion}a and\ \ref{fig:thin dispersion}b respectively show the calculated real part of the in-plane wavevector $\beta_{R}$ and the propagation length of SPP $\Lambda_\text{SPP}$, considering the size effect of permittivity of 10-nm-thick metal films on a glass substrate. For both interfaces of the metal film, $\beta_{R}$ aligns linearly with the light line except Pt, showing photon-like characteristics. Due to Pt's lower plasma frequency compared to other metals, it diverges from the light line in the high-frequency region, as explained by Eq.\ \eqref{Eq:approxi betar}. This less photon-like behavior of SPPs in the Pt film is due to the strong coupling between SPPs at both interfaces. Metals with a higher plasma frequency have a thinner penetration depth into the metal, leading to the decoupling of SPPs at thinner film thickness \cite{ordonez2023plasmon}. Therefore, for the same 10-nm thickness, SPPs propagating along the Pt films experience increased energy losses due to their stronger coupled nature, resulting in the shortest propagation length among selected metals, as shown in Fig.\ \ref{fig:thin dispersion}b. In the thin-film limit, the propagation length increases for metals with a higher plasma frequency and a lower damping coefficient, i.e., $\Lambda_\text{SPP}\propto\omega_{p}^{4}/\Gamma$, as shown in Eq.\ \eqref{Eq:approxi betai}. Therefore, the sequence of metals by longer propagation length is as follows: Al, Ag, Cu, Au, and Pt. Here, Cu and Au exhibit small difference in their propagation lengths. Specifically, while the fourth power of the plasma frequency for Au is 1.44 times greater than that for Cu ($(\omega_{p,\text{Au}}/\omega_{p,\text{Cu}})^{4}=1.44$), its damping coefficient is proportionally 1.48 times larger ($\Gamma_{\text{Au}}/\Gamma_{\text{Cu}}=1.48$). This ultimately results in the propagation length in Au being approximately 0.97 times shorter than that in Cu.

\begin{figure}[b!]%
\centering
\includegraphics[width=0.5\textwidth]{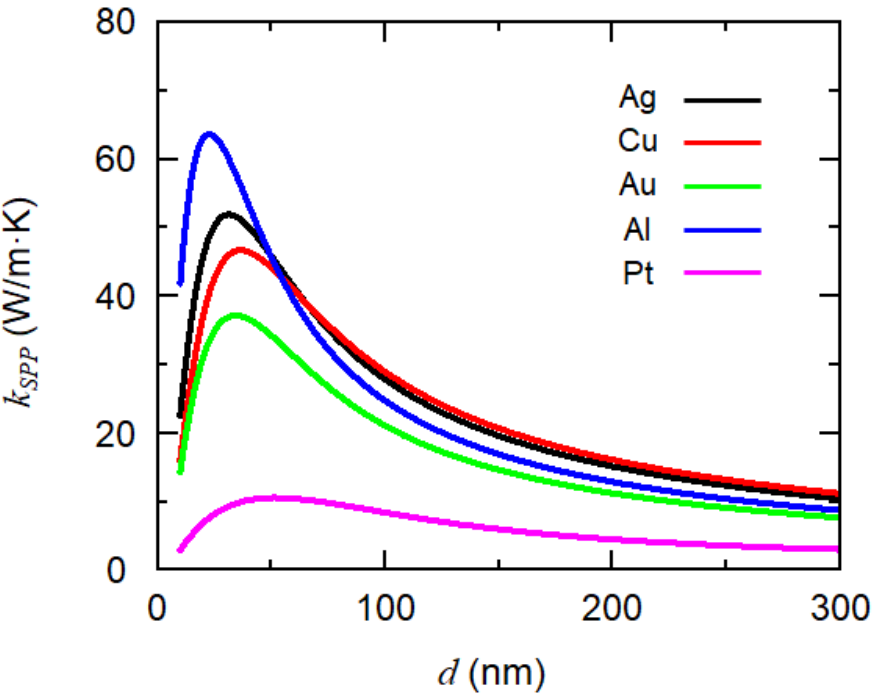}
\caption{Plasmon thermal conductivity of 10-cm-long metal films deposited on lossless glass substrate ($\varepsilon_{s}=3.6955$).}\label{fig:kSPP d}
\end{figure}

The dispersion relation for thin metal films with $L_m=10$ cm deposited on a glass substrate, as described in Eq.\ \eqref{Eq:dispersion}, was numerically solved. Using Eq.\ \eqref{Eq:kSPP}, the plasmon thermal conductivity as a function of film thickness was also calculated. To circumvent the non-local effect observed in extremely thin metallic layers, we focused on thicknesses exceeding 10 nm \cite{kim2023plasmon}. Figure\ \ref{fig:kSPP d} shows the calculated plasmon thermal conductivity for each metal as a function of their thickness. When $d_m < 20$ nm, the plasmon thermal conductivity increases as the film thickness increases, and the sequence of metals with high plasmon thermal conductivity is Al, Ag, Cu, Au, and Pt. This ordering mirrors the sequence observed for the longest propagation lengths shown in Fig.\ \ref{fig:thin dispersion}b. When the film thickness increases, the plasmon thermal conductivity first increases and reaches a peak, and then decreases. The sequence of metals with higher plasmon thermal conductivity changes if the film thickness increases to the optically thick limit. In the thick-film limit, the sequence of metals with high plasmon thermal conductivity changes to Cu, Ag, Al, Au, and Pt. This sequence is the same as the sequence of bulk metals with high electron thermal conductivity predicted from Wiedemann-Franz law, as shown in Table\ \ref{table 1}. 

In both the thin-film and thick-film limits, the plasmon thermal conductivity shares common variables: the Drude parameters and film thickness. In the thin-film limit, plasmon thermal conductivity is given by $k_\text{SPP,thin}\propto\omega_{p}^{4}d_m\Gamma^{-1}$, while in the thick-film limit, $k_\text{SPP,thick}\propto\omega_{p}^{2}d_m^{-1}\Gamma_{\infty}^{-1}$. In the intermediate regime (i.e., $\text{thin-film limit}<d_m<\text{thick-film limit}$), we hypothesized that the plasmon thermal conductivity can be assumed to vary with the exponents of $\omega_{p}$ and $d_m$. Consequently, the peak of the plasmon thermal conductivity for each metal is assumed to follow
\begin{equation} \label{Eq:kSPPmax}
	 k_{\text{SPP,peak}} \propto (\omega_{p})^{a}(d_{\text{SPP,opt}})^{b}(\Gamma)^{-1},
\end{equation}
where $d_{\text{SPP,opt}}$ denotes the optimal film thickness at which the plasmon thermal conductivity reaches its peak. Since the peak of the plasmon thermal conductivity exists at $d_m$ between the thin-film and thick-film limits, the exponents for $\omega_{p}$ and $d_m$ in the above equation should also satisfy values between those of the thin-film and thick-film limits; that is, the exponents $a$ and $b$ satisfy the ranges $2\leq a\leq 4$ and $-1\leq b\leq 1$, respectively. Note again that for finite-length metal films, the plasmon thermal conductivity is calculated using the effective propagation length. Consequently, the integral term with respect to angular frequency in Eqs.\ \eqref{Eq:kSPPthin} and\ \eqref{Eq:kSPPthick} becomes functions of Drude parameters, permittivity of substrates, and the thickness and length of the metal film (i.e., $\int_0^\infty g(\omega_{p},\Gamma_{\infty},\varepsilon_{s},d_{m},L_{m},\omega)\cdot\partial f_0/\partial T \, d\omega$). Since it is challenging to calculate the exponents for a finite-length film, to obtain the exponents, an infinite-length metal film where the integral term becomes constant should be considered first. For infinite-length metal film, we have $g(\omega_{p},\Gamma_{\infty},\varepsilon_{s},d_{m},L_{m},\omega)=1$ and $\int_0^\infty \partial f_0/\partial T \, d\omega=\text{constant}$. Table\ \ref{table 2} lists the optimal thickness and peak of the plasmon thermal conductivity for each infinite-length metal film. The process of calculating the exponents of the infinite-length metal film is as follows: First, determine the reference metal. We selected Ag as reference metal, because its thermal conductivity predicted by Drude parameters and Wiedemann-Franz law is the closest to  the value reported in the literature, as shown in Table.\ \ref{table 1}. Second, divide $k_{\text{SPP,peak}}$ and $d_{\text{SPP,opt}}$ for each metal by the values of the reference metal, i.e., $k_{\text{SPP,peak}}/k_{\text{SPP,peak,Ag}}=(\omega_{p}/\omega_{p,\text{Ag}})^{a}(d_{\text{SPP,opt}}/d_{\text{SPP,opt,Ag}})^{b}(\Gamma/\Gamma_{\text{Ag}})^{-1}$. In this way, the integral terms in both selected metal and the reference metal cancel out. Third, fit the exponents $a$ and $b$ of $k_{\text{SPP,peak}}/k_{\text{SPP,peak,Ag}}$ for all metals except Ag using the least-square method, which leads to $a=2.64$ and $b=-0.39$. Therefore, metals with a higher plasma frequency, thinner optimal film thickness, and lower damping coefficient exhibit higher peak value in the plasmon thermal conductivity (i.e., maximum plasmon thermal conductivity).

\begin{table}[t!]
\centering
\caption{Peak of the plasmon thermal conductivity and optimal film thickness of various thin metal films deposited on a lossless glass substrate. The damping coefficient of Drude model is estimated at the optimal thickness.}
\label{table 2}
\begin{tabular}{@{\hspace{5pt}}c@{\hspace{7pt}}|@{\hspace{5pt}}cc@{\hspace{5pt}}c}
\hline \hline
Metal & $d_{\text{SPP,opt}}$ (nm) & $\Gamma$ (cm$^{-1}$) & $k_{\text{SPP,peak}}$ (W/m$\cdot$K) \\ \hline \hline
Ag & 58.6 & 276.2 & 607.2 \\
Cu & 72.6 & 157.2 & 577.5 \\
Au & 58.3 & 377.8 & 335.9 \\
Al & 39.6 & 651.4 & 649.4 \\
Pt & 74.4 & 651.1 & 53.8 \\ \hline \hline
\end{tabular}%
\end{table}

Note that the plasma frequency and damping coefficient simultaneously affect both the peak of the plasmon thermal conductivity and the optimal film thickness. For a given superstrate and substrate, $d_{\text{SPP,opt}}$ at which the plasmon thermal conductivity exhibits its peak should be determined by the thickness-dependent permittivity of metal, i.e., $d_{\text{SPP,opt}}=f(\omega_{p},\Gamma_{\infty},v_{f})$. As listed in Table\ \ref{table 2}, $d_{\text{SPP,opt}}$ is generally thinner for metals with higher plasma frequencies and damping coefficients. However, the variation of $d_{\text{SPP,opt}}$ across various metals is smaller compared to the damping coefficient. Consequently, metals with a lower damping coefficient cause an increase in $k_{\text{SPP,peak}}$. For instance, Cu and Au, which have similar plasma frequencies, exhibit an optimal thickness ratio of 0.92, but their damping coefficient ratio is 2.40, i.e., $(d_{\text{SPP,opt,Cu}}/d_{\text{SPP,opt,Au}})^{b}=0.92$ and $(\Gamma_{\text{Cu}}/\Gamma_{\text{Au}})^{-1}=2.40$. This indicates that the effect on $k_{\text{SPP,peak}}$ is more dominated by the lower damping coefficient than by the thinner optimal thickness. Ultimately, metals with a higher plasma frequency and a lower damping coefficient possess a larger $k_{\text{SPP,peak}}$, thereby achieving the maximum plasmon thermal conductivity among metals. This effect of the Drude parameters on the maximum plasmon thermal conductivity is similarly observed in finite-length films. As seen in Fig.\ \ref{fig:kSPP d}, the sequence of metals with the larger $k_{\text{SPP,peak}}$ is Al, Ag, Cu, Au, and Pt. This sequence is consistent with the sequence of metals for larger $k_{\text{SPP,peak}}$ in infinite-length films, as listed in Table.\ \ref{table 2}. Among the five metals, Al film exhibits the maximum plasmon thermal conductivity. Further investigation into the Drude parameters of metals reveals that, among existing metals, Al possesses the highest plasma frequency and a relatively small damping coefficient, which results in the maximum plasmon thermal conductivity.

\begin{figure}[t!]%
\centering
\includegraphics[width=0.5\textwidth]{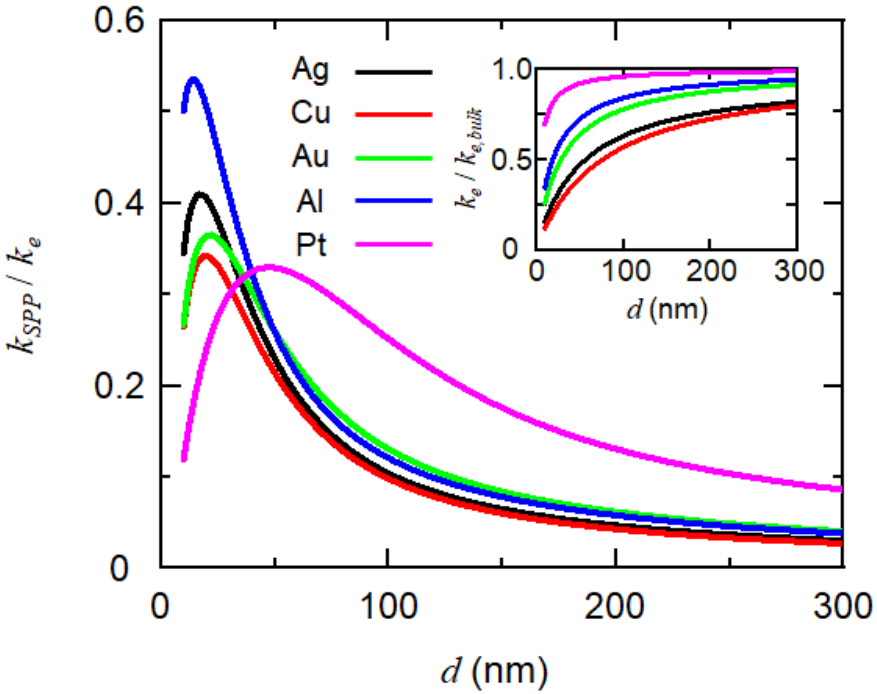}
\caption{The ratio of enhanced thermal conductivity due to SPPs compared to electron-mediated thermal conductivity in 10-cm-length thin metal films. The inset shows the normalized electron thermal conductivity for various metals calculated using the Wiedemann-Franz law, considering the classical size effect.}\label{fig:kSPP ke}
\end{figure}

The calculated $k_\text{SPP}/k_{e}$ is shown in Fig.\ \ref{fig:kSPP ke}, considering the thickness-dependent electron thermal conductivity at 300 K. Interestingly, it is observed that the optimal film thickness where this enhancement ratio peaks ($d_{\text{r,opt}}$) is 5\% to 35\% thinner than the optimal thickness for the peak of the plasmon thermal conductivity ($d_{\text{SPP,opt}}$). This is because while plasmon thermal conductivity attains its peak and has a gradual slope near the peak, the electron thermal conductivity rapidly decreases due to boundary scattering as film thickness decreases. Considering that the mean free path of electrons in bulk metals is tens of nanometers \cite{gall2016electron}, metal films with thickness in the tens of nanometers range cause sharp reduction in electron thermal conductivity, as shown in the inset of Fig.\ \ref{fig:kSPP ke}. To be specific, the electron thermal conductivity of Ag is 26\% of its bulk value at $d = 20$ nm, but it dramatically decreases to 15\% at $d = 10$ nm.

The peak of $k_\text{SPP}/k_{e}$ for each metal film can also be represented by a simple expression of $k_{\text{SPP}}/k_{e}\propto(\omega_{p})^{c}(d_{\text{r,opt}})^{d}$ because the electron thermal conductivity is given by $k_{e}\propto\omega_{p}^2/\Gamma$. Note that, unlike Eq.\ \eqref{Eq:kSPPmax}, the damping coefficient is canceled out in $k_{\text{SPP}}/k_{e}$. Since $d_{\text{r,opt}}<d_{\text{SPP,opt}}$, the exponent $c$ for plasma frequency should satisfy $a-2=0.64\leq c\leq2$. For the same reason, the exponent $d$ for film thickness should meet $b=-0.39\leq d\leq1$. These exponents $c$ and $d$ were fitted by the same method as before in Eq.\ \eqref{Eq:kSPPmax} with Ag as the reference metal, i.e., $(k_{\text{SPP}}/k_{e})/(k_{\text{SPP,Ag}}/k_{e,\text{Ag}})=(\omega_{p}/\omega_{p,\text{Ag}})^{c}(d_{\text{r,opt}}/d_{\text{r,opt,Ag}})^{d}$, resulting in $c=1.28$ and $d=0.17$. The difference in $d_{\text{r,opt}}$ for each metal is not significant, and since the exponent $d$ is small, the effect of $d_{\text{r,opt}}$ on the peak of $k_\text{SPP}/k_{e}$ is very small. Specifically, for Al and Pt films deposited on a lossless glass substrate, $d_{\text{r,opt}}$ is 29.6 nm for Al and 70.9 nm for Pt, leading to $(d_{\text{r,opt,Pt}}/d_{\text{r,opt,Al}})^{d}=1.16$. In contrast, $(\omega_{p,\text{Al}}/\omega_{p,\text{Pt}})^{c}=2.92$. Therefore, metals with higher plasma frequencies exhibit the largest value of $k_\text{SPP}/k_{e}$ (i.e., maximum enhancement ratio of thermal conductivity). As shown in Fig.\ \ref{fig:kSPP ke}, for instance, $k_\text{SPP}/k_{e}$ is larger in the sequence of metals with higher plasma frequency (i.e., Al, Ag, Au, Cu, and Pt). For a 10-cm-long and 14-nm-thick Al film deposited on a lossless glass substrate, the contribution of long-range SPPs is about 53.5\% of the electron counterpart. Analysis of $k_\text{SPP,peak}$ itself as well as $k_\text{SPP}/k_{e}$ commonly indicates that metals with a higher plasma frequency exhibit superior in-plane heat conduction via SPPs. This suggests that when applying a thin metallic film as a heat spreader, the priority should be given to metals with a high plasma frequency. If there are multiple candidates, then choose the metal with the lowest damping coefficient among them.

\begin{figure}[t!]%
\centering
\includegraphics[width=0.5\textwidth]{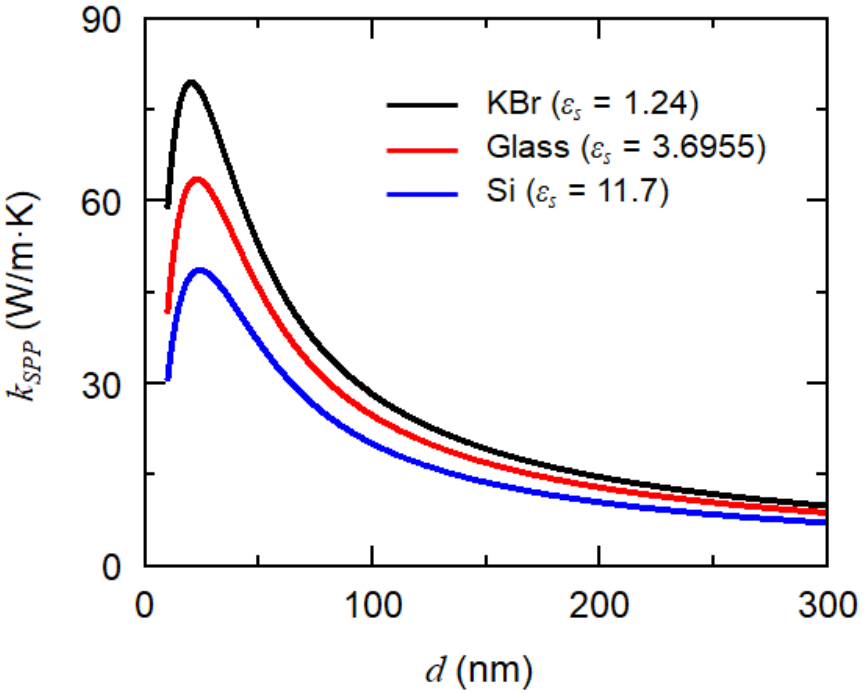}
\caption{Plasmon thermal conductivity of 10-cm-length metal films deposited on various substrates.}\label{fig:kSPP sub}
\end{figure}

The plasmon thermal conductivity of 10-cm-long Al thin film on various lossless substrates, such as KBr ($\varepsilon_{s}=1.24$) \cite{ordonez2013anomalous}, glass ($\varepsilon_{s}=3.6955$), and Si ($\varepsilon_{s}=11.7$) \cite{palik1998handbook}, is now considered, as shown in Fig.\ \ref{fig:kSPP sub}. A clear trend emerges: as the permittivity of the substrate decreases, the plasmon thermal conductivity consistently increases, across both the thin-film and thick-film limits. This relationship is rooted in the inverse proportionality between the substrate permittivity and the propagation length of SPPs, as detailed in Eq.\ \eqref{Eq:approxi betai}. Consequently, a lower substrate permittivity leads to the enhanced plasmon thermal conductivity in the thin-film limit, as formulated in Eq.\ \eqref{Eq:kSPPthin}. Additionally, in the thick-film limit, a lower substrate permittivity also increases the propagation of SPPs \cite{ordonez2023plasmon}, enhancing the plasmon thermal conductivity, as shown in Eq.\ \eqref{Eq:kSPPthick}. That is, in both thin-film and thick-film limits, the permittivity of substrate is inversely proportional to the plasmon thermal conductivity. Interestingly, this inverse proportionality between substrate permittivity and plasmon thermal conductivity is not confined to just these limits; it persists even for film thicknesses in the intermediate regime. Therefore, for a given metal film, the peak of the plasmon thermal conductivity increases as the permittivity of the substrate decreases, achieving the maximum plasmon thermal conductivity. Additionally, the peak of $k_\text{SPP}/k_{e}$ increases with lower substrate permittivity. For example, the peak of $k_\text{SPP}/k_{e}$ in Al film on a KBr substrate increases to 0.72, which is higher than that on a glass substrate. Choosing metals with high plasma frequencies and substrates with low permittivity indicates that plasmon thermal conductivity can more effectively compensate for the classical size effect of thin film thermal conductivity.

Finally, we extend our study to realistic situations when the superstrate is fixed as air while losses are included in the substrate. In the case of thin metal films deposited on lossy glass substrates, Al also turns out to be the best material leading to the maximum $k_\text{SPP,peak}$. Specifically, a 10-cm-long Al thin film deposited on the lossy glass substrate shows $k_\text{SPP}/k_{e}$ of 0.38 at the thickness of 12 nm. This enhancement ratio of thermal conductivity predominantly occurred due to the heat transfer of SPPs at the air/metal interface. This is because the lossy substrate absorbs the energy of SPPs, shortening their propagation length \cite{kim2023boosting, kim2023plasmon}.  Correspondingly, the contribution of heat transfer of SPPs at the metal/substrate interface to plasmon thermal conductivity becomes negligible. Therefore, the plasmon thermal conductivity of thin metal films deposited on lossy substrates is determined mainly by the behavior of SPPs at the air/metal interface. This observation leads to an intriguing conclusion: the established effect of permittivity of media on the plasmon thermal conductivity, which was initially formulated considering lossless media and based on the permittivity of the metal and substrate, remarkably holds true even in the cases with lossy substrates.

\section{Conclusions}
In this study, we derived an analytical expression for the plasmon thermal conductivity in both thin-film and thick-film limits as a function of metal film thickness, thickness-dependent metal permittivity, and substrate permittivity. Based on the Wiedemann-Franz law and the Drude model, it was shown that the plasmon thermal conductivity is proportional to the electron thermal conductivity. We also showed that metals with a high plasma frequency and a low damping coefficient can exhibit the maximum plasmon thermal conductivity as well as the maximum enhancement ratio of thermal conductivity. For instance, a 14-nm-thick and 10-cm-long Al film on a lossless glass substrate can achieve a 53.5\% enhancement in thermal conductivity compared to the electron thermal conductivity. These findings are crucial for optimizing the plasmon thermal conductivity of a thin metallic heat spreader that can be useful for advanced thermal management in electronic devices.

\begin{acknowledgments}
This research is supported by the Basic Science Research Program (NRF-20191A2C2003605) through the National Research Foundation of Korea (NRF) funded by the Ministry of Science and ICT.

\end{acknowledgments}


\bibliography{Refs.bib}

\providecommand{\noopsort}[1]{}\providecommand{\singleletter}[1]{#1}%
\begin{thebibliography}{38}%
\makeatletter
\providecommand \@ifxundefined [1]{%
 \@ifx{#1\undefined}
}%
\providecommand \@ifnum [1]{%
 \ifnum #1\expandafter \@firstoftwo
 \else \expandafter \@secondoftwo
 \fi
}%
\providecommand \@ifx [1]{%
 \ifx #1\expandafter \@firstoftwo
 \else \expandafter \@secondoftwo
 \fi
}%
\providecommand \natexlab [1]{#1}%
\providecommand \enquote  [1]{``#1''}%
\providecommand \bibnamefont  [1]{#1}%
\providecommand \bibfnamefont [1]{#1}%
\providecommand \citenamefont [1]{#1}%
\providecommand \href@noop [0]{\@secondoftwo}%
\providecommand \href [0]{\begingroup \@sanitize@url \@href}%
\providecommand \@href[1]{\@@startlink{#1}\@@href}%
\providecommand \@@href[1]{\endgroup#1\@@endlink}%
\providecommand \@sanitize@url [0]{\catcode `\\12\catcode `\$12\catcode
  `\&12\catcode `\#12\catcode `\^12\catcode `\_12\catcode `\%12\relax}%
\providecommand \@@startlink[1]{}%
\providecommand \@@endlink[0]{}%
\providecommand \url  [0]{\begingroup\@sanitize@url \@url }%
\providecommand \@url [1]{\endgroup\@href {#1}{\urlprefix }}%
\providecommand \urlprefix  [0]{URL }%
\providecommand \Eprint [0]{\href }%
\providecommand \doibase [0]{https://doi.org/}%
\providecommand \selectlanguage [0]{\@gobble}%
\providecommand \bibinfo  [0]{\@secondoftwo}%
\providecommand \bibfield  [0]{\@secondoftwo}%
\providecommand \translation [1]{[#1]}%
\providecommand \BibitemOpen [0]{}%
\providecommand \bibitemStop [0]{}%
\providecommand \bibitemNoStop [0]{.\EOS\space}%
\providecommand \EOS [0]{\spacefactor3000\relax}%
\providecommand \BibitemShut  [1]{\csname bibitem#1\endcsname}%
\let\auto@bib@innerbib\@empty
\bibitem [{\citenamefont {Chen}(2005)}]{chen2005nanoscale}%
  \BibitemOpen
  \bibfield  {author} {\bibinfo {author} {\bibfnamefont {G.}~\bibnamefont
  {Chen}},\ }\href@noop {} {\emph {\bibinfo {title} {Nanoscale Energy Transport
  and Conversion: A Parallel Treatment of Electrons, Molecules, Phonons, and
  Photons}}}\ (\bibinfo  {publisher} {Oxford university press},\ \bibinfo
  {year} {2005})\BibitemShut {NoStop}%
\bibitem [{\citenamefont {Zhang}\ \emph {et~al.}(2007)\citenamefont {Zhang},
  \citenamefont {Zhang},\ and\ \citenamefont {Luby}}]{zhang2007nano}%
  \BibitemOpen
  \bibfield  {author} {\bibinfo {author} {\bibfnamefont {Z.~M.}\ \bibnamefont
  {Zhang}}, \bibinfo {author} {\bibfnamefont {Z.~M.}\ \bibnamefont {Zhang}},\
  and\ \bibinfo {author} {\bibnamefont {Luby}},\ }\href@noop {} {\emph
  {\bibinfo {title} {Nano/Microscale Heat Transfer}}},\ Vol.\ \bibinfo {volume}
  {410}\ (\bibinfo  {publisher} {Springer},\ \bibinfo {year}
  {2007})\BibitemShut {NoStop}%
\bibitem [{\citenamefont {Feng}\ \emph {et~al.}(2009)\citenamefont {Feng},
  \citenamefont {Li},\ and\ \citenamefont {Zhang}}]{feng2009prediction}%
  \BibitemOpen
  \bibfield  {author} {\bibinfo {author} {\bibfnamefont {B.}~\bibnamefont
  {Feng}}, \bibinfo {author} {\bibfnamefont {Z.}~\bibnamefont {Li}},\ and\
  \bibinfo {author} {\bibfnamefont {X.}~\bibnamefont {Zhang}},\ }\href@noop {}
  {\bibfield  {journal} {\bibinfo  {journal} {Thin Solid Films}\ }\textbf
  {\bibinfo {volume} {517}},\ \bibinfo {pages} {2803} (\bibinfo {year}
  {2009})}\BibitemShut {NoStop}%
\bibitem [{\citenamefont {Jeong}\ \emph {et~al.}(2012)\citenamefont {Jeong},
  \citenamefont {Datta},\ and\ \citenamefont {Lundstrom}}]{jeong2012thermal}%
  \BibitemOpen
  \bibfield  {author} {\bibinfo {author} {\bibfnamefont {C.}~\bibnamefont
  {Jeong}}, \bibinfo {author} {\bibfnamefont {S.}~\bibnamefont {Datta}},\ and\
  \bibinfo {author} {\bibfnamefont {M.}~\bibnamefont {Lundstrom}},\ }\href@noop
  {} {\bibfield  {journal} {\bibinfo  {journal} {Journal of Applied Physics}\
  }\textbf {\bibinfo {volume} {111}},\ \bibinfo {pages} {093708} (\bibinfo
  {year} {2012})}\BibitemShut {NoStop}%
\bibitem [{\citenamefont {Wang}\ and\ \citenamefont
  {Huang}(2014)}]{wang2014computational}%
  \BibitemOpen
  \bibfield  {author} {\bibinfo {author} {\bibfnamefont {X.}~\bibnamefont
  {Wang}}\ and\ \bibinfo {author} {\bibfnamefont {B.}~\bibnamefont {Huang}},\
  }\href@noop {} {\bibfield  {journal} {\bibinfo  {journal} {Scientific
  Reports}\ }\textbf {\bibinfo {volume} {4}},\ \bibinfo {pages} {6399}
  (\bibinfo {year} {2014})}\BibitemShut {NoStop}%
\bibitem [{\citenamefont {Moore}\ and\ \citenamefont
  {Shi}(2014)}]{moore2014emerging}%
  \BibitemOpen
  \bibfield  {author} {\bibinfo {author} {\bibfnamefont {A.~L.}\ \bibnamefont
  {Moore}}\ and\ \bibinfo {author} {\bibfnamefont {L.}~\bibnamefont {Shi}},\
  }\href@noop {} {\bibfield  {journal} {\bibinfo  {journal} {Materials Today}\
  }\textbf {\bibinfo {volume} {17}},\ \bibinfo {pages} {163} (\bibinfo {year}
  {2014})}\BibitemShut {NoStop}%
\bibitem [{\citenamefont {Zheng}\ \emph {et~al.}(2021)\citenamefont {Zheng},
  \citenamefont {Hao}, \citenamefont {Miao}, \citenamefont {Schaadt},\ and\
  \citenamefont {Dames}}]{zheng2021advances}%
  \BibitemOpen
  \bibfield  {author} {\bibinfo {author} {\bibfnamefont {Q.}~\bibnamefont
  {Zheng}}, \bibinfo {author} {\bibfnamefont {M.}~\bibnamefont {Hao}}, \bibinfo
  {author} {\bibfnamefont {R.}~\bibnamefont {Miao}}, \bibinfo {author}
  {\bibfnamefont {J.}~\bibnamefont {Schaadt}},\ and\ \bibinfo {author}
  {\bibfnamefont {C.}~\bibnamefont {Dames}},\ }\href@noop {} {\bibfield
  {journal} {\bibinfo  {journal} {Progress in Energy}\ }\textbf {\bibinfo
  {volume} {3}},\ \bibinfo {pages} {012002} (\bibinfo {year}
  {2021})}\BibitemShut {NoStop}%
\bibitem [{\citenamefont {Zhou}\ \emph {et~al.}(2017)\citenamefont {Zhou},
  \citenamefont {Ramaneti}, \citenamefont {Anaya}, \citenamefont {Korneychuk},
  \citenamefont {Derluyn}, \citenamefont {Sun}, \citenamefont {Pomeroy},
  \citenamefont {Verbeeck}, \citenamefont {Haenen},\ and\ \citenamefont
  {Kuball}}]{zhou2017thermal}%
  \BibitemOpen
  \bibfield  {author} {\bibinfo {author} {\bibfnamefont {Y.}~\bibnamefont
  {Zhou}}, \bibinfo {author} {\bibfnamefont {R.}~\bibnamefont {Ramaneti}},
  \bibinfo {author} {\bibfnamefont {J.}~\bibnamefont {Anaya}}, \bibinfo
  {author} {\bibfnamefont {S.}~\bibnamefont {Korneychuk}}, \bibinfo {author}
  {\bibfnamefont {J.}~\bibnamefont {Derluyn}}, \bibinfo {author} {\bibfnamefont
  {H.}~\bibnamefont {Sun}}, \bibinfo {author} {\bibfnamefont {J.}~\bibnamefont
  {Pomeroy}}, \bibinfo {author} {\bibfnamefont {J.}~\bibnamefont {Verbeeck}},
  \bibinfo {author} {\bibfnamefont {K.}~\bibnamefont {Haenen}},\ and\ \bibinfo
  {author} {\bibfnamefont {M.}~\bibnamefont {Kuball}},\ }\href@noop {}
  {\bibfield  {journal} {\bibinfo  {journal} {Applied Physics Letters}\
  }\textbf {\bibinfo {volume} {111}},\ \bibinfo {pages} {041901} (\bibinfo
  {year} {2017})}\BibitemShut {NoStop}%
\bibitem [{\citenamefont {Yan}\ \emph {et~al.}(2012)\citenamefont {Yan},
  \citenamefont {Liu}, \citenamefont {Khan},\ and\ \citenamefont
  {Balandin}}]{yan2012graphene}%
  \BibitemOpen
  \bibfield  {author} {\bibinfo {author} {\bibfnamefont {Z.}~\bibnamefont
  {Yan}}, \bibinfo {author} {\bibfnamefont {G.}~\bibnamefont {Liu}}, \bibinfo
  {author} {\bibfnamefont {J.~M.}\ \bibnamefont {Khan}},\ and\ \bibinfo
  {author} {\bibfnamefont {A.~A.}\ \bibnamefont {Balandin}},\ }\href@noop {}
  {\bibfield  {journal} {\bibinfo  {journal} {Nature Communications}\ }\textbf
  {\bibinfo {volume} {3}},\ \bibinfo {pages} {827} (\bibinfo {year}
  {2012})}\BibitemShut {NoStop}%
\bibitem [{\citenamefont {Bae}\ \emph {et~al.}(2014)\citenamefont {Bae},
  \citenamefont {Shabani}, \citenamefont {Lee}, \citenamefont {Baeck},
  \citenamefont {Cho},\ and\ \citenamefont {Ahn}}]{bae2014graphene}%
  \BibitemOpen
  \bibfield  {author} {\bibinfo {author} {\bibfnamefont {S.-H.}\ \bibnamefont
  {Bae}}, \bibinfo {author} {\bibfnamefont {R.}~\bibnamefont {Shabani}},
  \bibinfo {author} {\bibfnamefont {J.-B.}\ \bibnamefont {Lee}}, \bibinfo
  {author} {\bibfnamefont {S.-J.}\ \bibnamefont {Baeck}}, \bibinfo {author}
  {\bibfnamefont {H.~J.}\ \bibnamefont {Cho}},\ and\ \bibinfo {author}
  {\bibfnamefont {J.-H.}\ \bibnamefont {Ahn}},\ }\href@noop {} {\bibfield
  {journal} {\bibinfo  {journal} {IEEE Transactions on Electron Devices}\
  }\textbf {\bibinfo {volume} {61}},\ \bibinfo {pages} {4171} (\bibinfo {year}
  {2014})}\BibitemShut {NoStop}%
\bibitem [{\citenamefont {Jeon}\ \emph {et~al.}(2021)\citenamefont {Jeon},
  \citenamefont {Lim}, \citenamefont {Bae}, \citenamefont {Kadirov},
  \citenamefont {Choi},\ and\ \citenamefont {Lee}}]{jeon2021suppression}%
  \BibitemOpen
  \bibfield  {author} {\bibinfo {author} {\bibfnamefont {D.}~\bibnamefont
  {Jeon}}, \bibinfo {author} {\bibfnamefont {J.}~\bibnamefont {Lim}}, \bibinfo
  {author} {\bibfnamefont {J.}~\bibnamefont {Bae}}, \bibinfo {author}
  {\bibfnamefont {A.}~\bibnamefont {Kadirov}}, \bibinfo {author} {\bibfnamefont
  {Y.}~\bibnamefont {Choi}},\ and\ \bibinfo {author} {\bibfnamefont
  {S.}~\bibnamefont {Lee}},\ }\href@noop {} {\bibfield  {journal} {\bibinfo
  {journal} {Applied Surface Science}\ }\textbf {\bibinfo {volume} {543}},\
  \bibinfo {pages} {148801} (\bibinfo {year} {2021})}\BibitemShut {NoStop}%
\bibitem [{\citenamefont {Song}\ \emph {et~al.}(2018)\citenamefont {Song},
  \citenamefont {Liu}, \citenamefont {Liu}, \citenamefont {Wu}, \citenamefont
  {Cheng},\ and\ \citenamefont {Kang}}]{song2018two}%
  \BibitemOpen
  \bibfield  {author} {\bibinfo {author} {\bibfnamefont {H.}~\bibnamefont
  {Song}}, \bibinfo {author} {\bibfnamefont {J.}~\bibnamefont {Liu}}, \bibinfo
  {author} {\bibfnamefont {B.}~\bibnamefont {Liu}}, \bibinfo {author}
  {\bibfnamefont {J.}~\bibnamefont {Wu}}, \bibinfo {author} {\bibfnamefont
  {H.-M.}\ \bibnamefont {Cheng}},\ and\ \bibinfo {author} {\bibfnamefont
  {F.}~\bibnamefont {Kang}},\ }\href@noop {} {\bibfield  {journal} {\bibinfo
  {journal} {Joule}\ }\textbf {\bibinfo {volume} {2}},\ \bibinfo {pages} {442}
  (\bibinfo {year} {2018})}\BibitemShut {NoStop}%
\bibitem [{\citenamefont {Chen}\ \emph {et~al.}(2005)\citenamefont {Chen},
  \citenamefont {Narayanaswamy},\ and\ \citenamefont {Chen}}]{chen2005surface}%
  \BibitemOpen
  \bibfield  {author} {\bibinfo {author} {\bibfnamefont {D.-Z.~A.}\
  \bibnamefont {Chen}}, \bibinfo {author} {\bibfnamefont {A.}~\bibnamefont
  {Narayanaswamy}},\ and\ \bibinfo {author} {\bibfnamefont {G.}~\bibnamefont
  {Chen}},\ }\href@noop {} {\bibfield  {journal} {\bibinfo  {journal} {Physical
  Review B}\ }\textbf {\bibinfo {volume} {72}},\ \bibinfo {pages} {155435}
  (\bibinfo {year} {2005})}\BibitemShut {NoStop}%
\bibitem [{\citenamefont {Tranchant}\ \emph {et~al.}(2019)\citenamefont
  {Tranchant}, \citenamefont {Hamamura}, \citenamefont {Ordonez-Miranda},
  \citenamefont {Yabuki}, \citenamefont {Vega-Flick}, \citenamefont
  {Cervantes-Alvarez}, \citenamefont {Alvarado-Gil}, \citenamefont {Volz},\
  and\ \citenamefont {Miyazaki}}]{tranchant2019two}%
  \BibitemOpen
  \bibfield  {author} {\bibinfo {author} {\bibfnamefont {L.}~\bibnamefont
  {Tranchant}}, \bibinfo {author} {\bibfnamefont {S.}~\bibnamefont {Hamamura}},
  \bibinfo {author} {\bibfnamefont {J.}~\bibnamefont {Ordonez-Miranda}},
  \bibinfo {author} {\bibfnamefont {T.}~\bibnamefont {Yabuki}}, \bibinfo
  {author} {\bibfnamefont {A.}~\bibnamefont {Vega-Flick}}, \bibinfo {author}
  {\bibfnamefont {F.}~\bibnamefont {Cervantes-Alvarez}}, \bibinfo {author}
  {\bibfnamefont {J.~J.}\ \bibnamefont {Alvarado-Gil}}, \bibinfo {author}
  {\bibfnamefont {S.}~\bibnamefont {Volz}},\ and\ \bibinfo {author}
  {\bibfnamefont {K.}~\bibnamefont {Miyazaki}},\ }\href@noop {} {\bibfield
  {journal} {\bibinfo  {journal} {Nano Letters}\ }\textbf {\bibinfo {volume}
  {19}},\ \bibinfo {pages} {6924} (\bibinfo {year} {2019})}\BibitemShut
  {NoStop}%
\bibitem [{\citenamefont {Wu}\ \emph {et~al.}(2020)\citenamefont {Wu},
  \citenamefont {Ordonez-Miranda}, \citenamefont {Gluchko}, \citenamefont
  {Anufriev}, \citenamefont {Meneses}, \citenamefont {Del~Campo}, \citenamefont
  {Volz},\ and\ \citenamefont {Nomura}}]{wu2020enhanced}%
  \BibitemOpen
  \bibfield  {author} {\bibinfo {author} {\bibfnamefont {Y.}~\bibnamefont
  {Wu}}, \bibinfo {author} {\bibfnamefont {J.}~\bibnamefont {Ordonez-Miranda}},
  \bibinfo {author} {\bibfnamefont {S.}~\bibnamefont {Gluchko}}, \bibinfo
  {author} {\bibfnamefont {R.}~\bibnamefont {Anufriev}}, \bibinfo {author}
  {\bibfnamefont {D.~D.~S.}\ \bibnamefont {Meneses}}, \bibinfo {author}
  {\bibfnamefont {L.}~\bibnamefont {Del~Campo}}, \bibinfo {author}
  {\bibfnamefont {S.}~\bibnamefont {Volz}},\ and\ \bibinfo {author}
  {\bibfnamefont {M.}~\bibnamefont {Nomura}},\ }\href@noop {} {\bibfield
  {journal} {\bibinfo  {journal} {Science Advances}\ }\textbf {\bibinfo
  {volume} {6}},\ \bibinfo {pages} {eabb4461} (\bibinfo {year}
  {2020})}\BibitemShut {NoStop}%
\bibitem [{\citenamefont {Wu}\ \emph {et~al.}(2022)\citenamefont {Wu},
  \citenamefont {Ordonez-Miranda}, \citenamefont {Jalabert}, \citenamefont
  {Tachikawa}, \citenamefont {Anufriev}, \citenamefont {Fujita}, \citenamefont
  {Volz},\ and\ \citenamefont {Nomura}}]{wu2022observation}%
  \BibitemOpen
  \bibfield  {author} {\bibinfo {author} {\bibfnamefont {Y.}~\bibnamefont
  {Wu}}, \bibinfo {author} {\bibfnamefont {J.}~\bibnamefont {Ordonez-Miranda}},
  \bibinfo {author} {\bibfnamefont {L.}~\bibnamefont {Jalabert}}, \bibinfo
  {author} {\bibfnamefont {S.}~\bibnamefont {Tachikawa}}, \bibinfo {author}
  {\bibfnamefont {R.}~\bibnamefont {Anufriev}}, \bibinfo {author}
  {\bibfnamefont {H.}~\bibnamefont {Fujita}}, \bibinfo {author} {\bibfnamefont
  {S.}~\bibnamefont {Volz}},\ and\ \bibinfo {author} {\bibfnamefont
  {M.}~\bibnamefont {Nomura}},\ }\href@noop {} {\bibfield  {journal} {\bibinfo
  {journal} {Applied Physics Letters}\ }\textbf {\bibinfo {volume} {121}},\
  \bibinfo {pages} {112203} (\bibinfo {year} {2022})}\BibitemShut {NoStop}%
\bibitem [{\citenamefont {Kim}\ \emph {et~al.}(2023{\natexlab{a}})\citenamefont
  {Kim}, \citenamefont {Choi}, \citenamefont {Cho}, \citenamefont {Lim},\ and\
  \citenamefont {Lee}}]{kim2023boosting}%
  \BibitemOpen
  \bibfield  {author} {\bibinfo {author} {\bibfnamefont {D.-m.}\ \bibnamefont
  {Kim}}, \bibinfo {author} {\bibfnamefont {S.}~\bibnamefont {Choi}}, \bibinfo
  {author} {\bibfnamefont {J.}~\bibnamefont {Cho}}, \bibinfo {author}
  {\bibfnamefont {M.}~\bibnamefont {Lim}},\ and\ \bibinfo {author}
  {\bibfnamefont {B.~J.}\ \bibnamefont {Lee}},\ }\href@noop {} {\bibfield
  {journal} {\bibinfo  {journal} {Physical Review Letters}\ }\textbf {\bibinfo
  {volume} {130}},\ \bibinfo {pages} {176302} (\bibinfo {year}
  {2023}{\natexlab{a}})}\BibitemShut {NoStop}%
\bibitem [{\citenamefont {Pan}\ \emph {et~al.}(2023)\citenamefont {Pan},
  \citenamefont {Lu}, \citenamefont {Li}, \citenamefont {McBride},
  \citenamefont {Juneja}, \citenamefont {Long}, \citenamefont {Lindsay},
  \citenamefont {Caldwell},\ and\ \citenamefont {Li}}]{pan2023remarkable}%
  \BibitemOpen
  \bibfield  {author} {\bibinfo {author} {\bibfnamefont {Z.}~\bibnamefont
  {Pan}}, \bibinfo {author} {\bibfnamefont {G.}~\bibnamefont {Lu}}, \bibinfo
  {author} {\bibfnamefont {X.}~\bibnamefont {Li}}, \bibinfo {author}
  {\bibfnamefont {J.~R.}\ \bibnamefont {McBride}}, \bibinfo {author}
  {\bibfnamefont {R.}~\bibnamefont {Juneja}}, \bibinfo {author} {\bibfnamefont
  {M.}~\bibnamefont {Long}}, \bibinfo {author} {\bibfnamefont {L.}~\bibnamefont
  {Lindsay}}, \bibinfo {author} {\bibfnamefont {J.~D.}\ \bibnamefont
  {Caldwell}},\ and\ \bibinfo {author} {\bibfnamefont {D.}~\bibnamefont {Li}},\
  }\href@noop {} {\bibfield  {journal} {\bibinfo  {journal} {Nature}\ }\textbf
  {\bibinfo {volume} {623}},\ \bibinfo {pages} {307} (\bibinfo {year}
  {2023})}\BibitemShut {NoStop}%
\bibitem [{\citenamefont {Pei}\ \emph {et~al.}(2023)\citenamefont {Pei},
  \citenamefont {Chen}, \citenamefont {Jeon}, \citenamefont {Liu},\ and\
  \citenamefont {Chen}}]{pei2023low}%
  \BibitemOpen
  \bibfield  {author} {\bibinfo {author} {\bibfnamefont {Y.}~\bibnamefont
  {Pei}}, \bibinfo {author} {\bibfnamefont {L.}~\bibnamefont {Chen}}, \bibinfo
  {author} {\bibfnamefont {W.}~\bibnamefont {Jeon}}, \bibinfo {author}
  {\bibfnamefont {Z.}~\bibnamefont {Liu}},\ and\ \bibinfo {author}
  {\bibfnamefont {R.}~\bibnamefont {Chen}},\ }\href@noop {} {\bibfield
  {journal} {\bibinfo  {journal} {Nature Communications}\ }\textbf {\bibinfo
  {volume} {14}},\ \bibinfo {pages} {8242} (\bibinfo {year}
  {2023})}\BibitemShut {NoStop}%
\bibitem [{\citenamefont {Ordonez-Miranda}\ \emph {et~al.}(2013)\citenamefont
  {Ordonez-Miranda}, \citenamefont {Tranchant}, \citenamefont {Tokunaga},
  \citenamefont {Kim}, \citenamefont {Palpant}, \citenamefont {Chalopin},
  \citenamefont {Antoni},\ and\ \citenamefont {Volz}}]{ordonez2013anomalous}%
  \BibitemOpen
  \bibfield  {author} {\bibinfo {author} {\bibfnamefont {J.}~\bibnamefont
  {Ordonez-Miranda}}, \bibinfo {author} {\bibfnamefont {L.}~\bibnamefont
  {Tranchant}}, \bibinfo {author} {\bibfnamefont {T.}~\bibnamefont {Tokunaga}},
  \bibinfo {author} {\bibfnamefont {B.}~\bibnamefont {Kim}}, \bibinfo {author}
  {\bibfnamefont {B.}~\bibnamefont {Palpant}}, \bibinfo {author} {\bibfnamefont
  {Y.}~\bibnamefont {Chalopin}}, \bibinfo {author} {\bibfnamefont
  {T.}~\bibnamefont {Antoni}},\ and\ \bibinfo {author} {\bibfnamefont
  {S.}~\bibnamefont {Volz}},\ }\href@noop {} {\bibfield  {journal} {\bibinfo
  {journal} {Journal of Applied Physics}\ }\textbf {\bibinfo {volume} {113}},\
  \bibinfo {pages} {084311} (\bibinfo {year} {2013})}\BibitemShut {NoStop}%
\bibitem [{\citenamefont {Ordonez-Miranda}\ \emph {et~al.}(2014)\citenamefont
  {Ordonez-Miranda}, \citenamefont {Tranchant}, \citenamefont {Chalopin},
  \citenamefont {Antoni},\ and\ \citenamefont {Volz}}]{ordonez2014thermal}%
  \BibitemOpen
  \bibfield  {author} {\bibinfo {author} {\bibfnamefont {J.}~\bibnamefont
  {Ordonez-Miranda}}, \bibinfo {author} {\bibfnamefont {L.}~\bibnamefont
  {Tranchant}}, \bibinfo {author} {\bibfnamefont {Y.}~\bibnamefont {Chalopin}},
  \bibinfo {author} {\bibfnamefont {T.}~\bibnamefont {Antoni}},\ and\ \bibinfo
  {author} {\bibfnamefont {S.}~\bibnamefont {Volz}},\ }\href@noop {} {\bibfield
   {journal} {\bibinfo  {journal} {Journal of Applied Physics}\ }\textbf
  {\bibinfo {volume} {115}},\ \bibinfo {pages} {054311} (\bibinfo {year}
  {2014})}\BibitemShut {NoStop}%
\bibitem [{\citenamefont {Ordonez-Miranda}\ \emph {et~al.}(2023)\citenamefont
  {Ordonez-Miranda}, \citenamefont {Kosevich}, \citenamefont {Lee},
  \citenamefont {Nomura},\ and\ \citenamefont {Volz}}]{ordonez2023plasmon}%
  \BibitemOpen
  \bibfield  {author} {\bibinfo {author} {\bibfnamefont {J.}~\bibnamefont
  {Ordonez-Miranda}}, \bibinfo {author} {\bibfnamefont {Y.~A.}\ \bibnamefont
  {Kosevich}}, \bibinfo {author} {\bibfnamefont {B.~J.}\ \bibnamefont {Lee}},
  \bibinfo {author} {\bibfnamefont {M.}~\bibnamefont {Nomura}},\ and\ \bibinfo
  {author} {\bibfnamefont {S.}~\bibnamefont {Volz}},\ }\href@noop {} {\bibfield
   {journal} {\bibinfo  {journal} {Physical Review Applied}\ }\textbf {\bibinfo
  {volume} {19}},\ \bibinfo {pages} {044046} (\bibinfo {year}
  {2023})}\BibitemShut {NoStop}%
\bibitem [{\citenamefont {Kim}\ \emph {et~al.}(2023{\natexlab{b}})\citenamefont
  {Kim}, \citenamefont {Nam},\ and\ \citenamefont {Lee}}]{kim2023plasmon}%
  \BibitemOpen
  \bibfield  {author} {\bibinfo {author} {\bibfnamefont {D.-m.}\ \bibnamefont
  {Kim}}, \bibinfo {author} {\bibfnamefont {J.}~\bibnamefont {Nam}},\ and\
  \bibinfo {author} {\bibfnamefont {B.~J.}\ \bibnamefont {Lee}},\ }\href@noop
  {} {\bibfield  {journal} {\bibinfo  {journal} {Physical Review B}\ }\textbf
  {\bibinfo {volume} {108}},\ \bibinfo {pages} {205418} (\bibinfo {year}
  {2023}{\natexlab{b}})}\BibitemShut {NoStop}%
\bibitem [{\citenamefont {Guo}\ \emph {et~al.}(2021)\citenamefont {Guo},
  \citenamefont {Tachikawa}, \citenamefont {Volz}, \citenamefont {Nomura},\
  and\ \citenamefont {Ordonez-Miranda}}]{guo2021quantum}%
  \BibitemOpen
  \bibfield  {author} {\bibinfo {author} {\bibfnamefont {Y.}~\bibnamefont
  {Guo}}, \bibinfo {author} {\bibfnamefont {S.}~\bibnamefont {Tachikawa}},
  \bibinfo {author} {\bibfnamefont {S.}~\bibnamefont {Volz}}, \bibinfo {author}
  {\bibfnamefont {M.}~\bibnamefont {Nomura}},\ and\ \bibinfo {author}
  {\bibfnamefont {J.}~\bibnamefont {Ordonez-Miranda}},\ }\href@noop {}
  {\bibfield  {journal} {\bibinfo  {journal} {Physical Review B}\ }\textbf
  {\bibinfo {volume} {104}},\ \bibinfo {pages} {L201407} (\bibinfo {year}
  {2021})}\BibitemShut {NoStop}%
\bibitem [{\citenamefont {Yeh}\ and\ \citenamefont
  {Shimabukuro}(2008)}]{yeh2008plasmon}%
  \BibitemOpen
  \bibfield  {author} {\bibinfo {author} {\bibfnamefont {C.}~\bibnamefont
  {Yeh}}\ and\ \bibinfo {author} {\bibfnamefont {F.}~\bibnamefont
  {Shimabukuro}},\ }\href@noop {} {\emph {\bibinfo {title} {The Essence of
  Dielectric Waveguides}}}\ (\bibinfo  {publisher} {Springer},\ \bibinfo {year}
  {2008})\BibitemShut {NoStop}%
\bibitem [{\citenamefont {Maier}\ \emph {et~al.}(2007)\citenamefont {Maier}
  \emph {et~al.}}]{maier2007plasmonics}%
  \BibitemOpen
  \bibfield  {author} {\bibinfo {author} {\bibfnamefont {S.~A.}\ \bibnamefont
  {Maier}} \emph {et~al.},\ }\href@noop {} {\emph {\bibinfo {title}
  {Plasmonics: Fundamentals and Applications}}},\ Vol.~\bibinfo {volume} {1}\
  (\bibinfo  {publisher} {Springer},\ \bibinfo {year} {2007})\BibitemShut
  {NoStop}%
\bibitem [{\citenamefont {Otanicar}\ \emph {et~al.}(2010)\citenamefont
  {Otanicar}, \citenamefont {Phelan}, \citenamefont {Prasher}, \citenamefont
  {Rosengarten},\ and\ \citenamefont {Taylor}}]{otanicar2010nanofluid}%
  \BibitemOpen
  \bibfield  {author} {\bibinfo {author} {\bibfnamefont {T.~P.}\ \bibnamefont
  {Otanicar}}, \bibinfo {author} {\bibfnamefont {P.~E.}\ \bibnamefont
  {Phelan}}, \bibinfo {author} {\bibfnamefont {R.~S.}\ \bibnamefont {Prasher}},
  \bibinfo {author} {\bibfnamefont {G.}~\bibnamefont {Rosengarten}},\ and\
  \bibinfo {author} {\bibfnamefont {R.~A.}\ \bibnamefont {Taylor}},\
  }\href@noop {} {\bibfield  {journal} {\bibinfo  {journal} {Journal of
  Renewable and Sustainable Energy}\ }\textbf {\bibinfo {volume} {2}},\
  \bibinfo {pages} {033102} (\bibinfo {year} {2010})}\BibitemShut {NoStop}%
\bibitem [{\citenamefont {Lee}\ \emph {et~al.}(2012)\citenamefont {Lee},
  \citenamefont {Park}, \citenamefont {Walsh},\ and\ \citenamefont
  {Xu}}]{lee2012radiative}%
  \BibitemOpen
  \bibfield  {author} {\bibinfo {author} {\bibfnamefont {B.~J.}\ \bibnamefont
  {Lee}}, \bibinfo {author} {\bibfnamefont {K.}~\bibnamefont {Park}}, \bibinfo
  {author} {\bibfnamefont {T.}~\bibnamefont {Walsh}},\ and\ \bibinfo {author}
  {\bibfnamefont {L.}~\bibnamefont {Xu}},\ }\href@noop {} {\bibfield  {journal}
  {\bibinfo  {journal} {Journal of Solar Energy Engineering}\ }\textbf
  {\bibinfo {volume} {134}},\ \bibinfo {pages} {021009} (\bibinfo {year}
  {2012})}\BibitemShut {NoStop}%
\bibitem [{\citenamefont {Avery}\ \emph {et~al.}(2015)\citenamefont {Avery},
  \citenamefont {Mason}, \citenamefont {Bassett}, \citenamefont {Wesenberg},\
  and\ \citenamefont {Zink}}]{avery2015thermal}%
  \BibitemOpen
  \bibfield  {author} {\bibinfo {author} {\bibfnamefont {A.}~\bibnamefont
  {Avery}}, \bibinfo {author} {\bibfnamefont {S.}~\bibnamefont {Mason}},
  \bibinfo {author} {\bibfnamefont {D.}~\bibnamefont {Bassett}}, \bibinfo
  {author} {\bibfnamefont {D.}~\bibnamefont {Wesenberg}},\ and\ \bibinfo
  {author} {\bibfnamefont {B.}~\bibnamefont {Zink}},\ }\href@noop {} {\bibfield
   {journal} {\bibinfo  {journal} {Physical Review B}\ }\textbf {\bibinfo
  {volume} {92}},\ \bibinfo {pages} {214410} (\bibinfo {year}
  {2015})}\BibitemShut {NoStop}%
\bibitem [{\citenamefont {Kittel}\ and\ \citenamefont
  {McEuen}(2018)}]{kittel2018introduction}%
  \BibitemOpen
  \bibfield  {author} {\bibinfo {author} {\bibfnamefont {C.}~\bibnamefont
  {Kittel}}\ and\ \bibinfo {author} {\bibfnamefont {P.}~\bibnamefont
  {McEuen}},\ }\href@noop {} {\emph {\bibinfo {title} {Introduction to Solid
  State Physics}}}\ (\bibinfo  {publisher} {John Wiley \& Sons},\ \bibinfo
  {year} {2018})\BibitemShut {NoStop}%
\bibitem [{\citenamefont {Ordal}\ \emph {et~al.}(1985)\citenamefont {Ordal},
  \citenamefont {Bell}, \citenamefont {Alexander}, \citenamefont {Long},\ and\
  \citenamefont {Querry}}]{ordal1985optical}%
  \BibitemOpen
  \bibfield  {author} {\bibinfo {author} {\bibfnamefont {M.~A.}\ \bibnamefont
  {Ordal}}, \bibinfo {author} {\bibfnamefont {R.~J.}\ \bibnamefont {Bell}},
  \bibinfo {author} {\bibfnamefont {R.~W.}\ \bibnamefont {Alexander}}, \bibinfo
  {author} {\bibfnamefont {L.~L.}\ \bibnamefont {Long}},\ and\ \bibinfo
  {author} {\bibfnamefont {M.~R.}\ \bibnamefont {Querry}},\ }\href@noop {}
  {\bibfield  {journal} {\bibinfo  {journal} {Applied Optics}\ }\textbf
  {\bibinfo {volume} {24}},\ \bibinfo {pages} {4493} (\bibinfo {year}
  {1985})}\BibitemShut {NoStop}%
\bibitem [{\citenamefont {Ordal}\ \emph {et~al.}(1988)\citenamefont {Ordal},
  \citenamefont {Bell}, \citenamefont {Alexander}, \citenamefont {Newquist},\
  and\ \citenamefont {Querry}}]{ordal1988optical}%
  \BibitemOpen
  \bibfield  {author} {\bibinfo {author} {\bibfnamefont {M.~A.}\ \bibnamefont
  {Ordal}}, \bibinfo {author} {\bibfnamefont {R.~J.}\ \bibnamefont {Bell}},
  \bibinfo {author} {\bibfnamefont {R.~W.}\ \bibnamefont {Alexander}}, \bibinfo
  {author} {\bibfnamefont {L.~A.}\ \bibnamefont {Newquist}},\ and\ \bibinfo
  {author} {\bibfnamefont {M.~R.}\ \bibnamefont {Querry}},\ }\href@noop {}
  {\bibfield  {journal} {\bibinfo  {journal} {Applied Optics}\ }\textbf
  {\bibinfo {volume} {27}},\ \bibinfo {pages} {1203} (\bibinfo {year}
  {1988})}\BibitemShut {NoStop}%
\bibitem [{\citenamefont {Palik}(1998)}]{palik1998handbook}%
  \BibitemOpen
  \bibfield  {author} {\bibinfo {author} {\bibfnamefont {E.~D.}\ \bibnamefont
  {Palik}},\ }\href@noop {} {\emph {\bibinfo {title} {Handbook of Optical
  Constants of Solids}}},\ Vol.~\bibinfo {volume} {1}\ (\bibinfo  {publisher}
  {Academic press},\ \bibinfo {year} {1998})\BibitemShut {NoStop}%
\bibitem [{\citenamefont {Gall}(2016)}]{gall2016electron}%
  \BibitemOpen
  \bibfield  {author} {\bibinfo {author} {\bibfnamefont {D.}~\bibnamefont
  {Gall}},\ }\href@noop {} {\bibfield  {journal} {\bibinfo  {journal} {Journal
  of Applied Physics}\ }\textbf {\bibinfo {volume} {119}},\ \bibinfo {pages}
  {085101} (\bibinfo {year} {2016})}\BibitemShut {NoStop}%
\bibitem [{\citenamefont {Bergman}(2011)}]{bergman2011fundamentals}%
  \BibitemOpen
  \bibfield  {author} {\bibinfo {author} {\bibfnamefont {T.~L.}\ \bibnamefont
  {Bergman}},\ }\href@noop {} {\emph {\bibinfo {title} {Fundamentals of Heat
  and Mass Transfer}}}\ (\bibinfo  {publisher} {John Wiley \& Sons},\ \bibinfo
  {year} {2011})\BibitemShut {NoStop}%
\bibitem [{\citenamefont {Dressel}\ and\ \citenamefont
  {Scheffler}(2006)}]{dressel2006verifying}%
  \BibitemOpen
  \bibfield  {author} {\bibinfo {author} {\bibfnamefont {M.}~\bibnamefont
  {Dressel}}\ and\ \bibinfo {author} {\bibfnamefont {M.}~\bibnamefont
  {Scheffler}},\ }\href@noop {} {\bibfield  {journal} {\bibinfo  {journal}
  {Annalen der Physik}\ }\textbf {\bibinfo {volume} {518}},\ \bibinfo {pages}
  {535} (\bibinfo {year} {2006})}\BibitemShut {NoStop}%
\bibitem [{\citenamefont {Tong}\ \emph {et~al.}(2019)\citenamefont {Tong},
  \citenamefont {Li}, \citenamefont {Ruan},\ and\ \citenamefont
  {Bao}}]{tong2019comprehensive}%
  \BibitemOpen
  \bibfield  {author} {\bibinfo {author} {\bibfnamefont {Z.}~\bibnamefont
  {Tong}}, \bibinfo {author} {\bibfnamefont {S.}~\bibnamefont {Li}}, \bibinfo
  {author} {\bibfnamefont {X.}~\bibnamefont {Ruan}},\ and\ \bibinfo {author}
  {\bibfnamefont {H.}~\bibnamefont {Bao}},\ }\href@noop {} {\bibfield
  {journal} {\bibinfo  {journal} {Physical Review B}\ }\textbf {\bibinfo
  {volume} {100}},\ \bibinfo {pages} {144306} (\bibinfo {year}
  {2019})}\BibitemShut {NoStop}%
\bibitem [{\citenamefont {Dutta}\ \emph {et~al.}(2017)\citenamefont {Dutta},
  \citenamefont {Sankaran}, \citenamefont {Moors}, \citenamefont {Pourtois},
  \citenamefont {Van~Elshocht}, \citenamefont {B{\"o}mmels}, \citenamefont
  {Vandervorst}, \citenamefont {T{\H{o}}kei},\ and\ \citenamefont
  {Adelmann}}]{dutta2017thickness}%
  \BibitemOpen
  \bibfield  {author} {\bibinfo {author} {\bibfnamefont {S.}~\bibnamefont
  {Dutta}}, \bibinfo {author} {\bibfnamefont {K.}~\bibnamefont {Sankaran}},
  \bibinfo {author} {\bibfnamefont {K.}~\bibnamefont {Moors}}, \bibinfo
  {author} {\bibfnamefont {G.}~\bibnamefont {Pourtois}}, \bibinfo {author}
  {\bibfnamefont {S.}~\bibnamefont {Van~Elshocht}}, \bibinfo {author}
  {\bibfnamefont {J.}~\bibnamefont {B{\"o}mmels}}, \bibinfo {author}
  {\bibfnamefont {W.}~\bibnamefont {Vandervorst}}, \bibinfo {author}
  {\bibfnamefont {Z.}~\bibnamefont {T{\H{o}}kei}},\ and\ \bibinfo {author}
  {\bibfnamefont {C.}~\bibnamefont {Adelmann}},\ }\href@noop {} {\bibfield
  {journal} {\bibinfo  {journal} {Journal of Applied Physics}\ }\textbf
  {\bibinfo {volume} {122}},\ \bibinfo {pages} {025107} (\bibinfo {year}
  {2017})}\BibitemShut {NoStop}%
\end{thebibliography}%

%

\end{document}